\documentclass{aa}
\usepackage{epsf}

\begin{document}
\sloppypar

   \thesaurus{06     
              (02.01.2;  
               02.09.1;  
	       08.02.3;  
	       08.14.1;  
               13.25.3;  
               13.25.5)} 
   \title{New class of low frequency QPOs: signature of
nuclear burning or accretion disk instabilities?}	

   \author{M. Revnivtsev \inst{1,2}, E. Churazov \inst{2,1}, 
	M. Gilfanov \inst{2,1}, R. Sunyaev \inst{2,1}}

   \offprints{revnivtsev@hea.iki.rssi.ru}

   \institute{Space Research Institute, Russian Academy of Sciences,
              Profsoyuznaya 84/32, 117810 Moscow, Russia 
	\and
	      Max-Planck-Institute f\"ur Astrophysik,
              Karl-Schwarzschild-Str. 1, D-85740 Garching bei M\"unchen,
              Germany,
            }
  \date{Received 6 November 2000/ Accepted 28 February 2001}

        \authorrunning{Revnivtsev et al.}
        \titlerunning{New class of low frequency QPOs: signature of
nuclear burning or disk instabilities?}
        
   \maketitle

   \begin{abstract}

We report the discovery of a new class of low frequency
quasi--periodic variations of the X--ray flux in the X-ray bursters
4U1608-52 and 
4U1636-536. We also report an occasional detection of a similar QPO in
Aql X-1. The QPOs, associated with flux variations at the level
of percents, are observed at a frequency of 7--9 $\cdot 10^{-3}$
Hz. While usually the relative amplitude of flux variations increases
with energy, the newly discovered QPOs are limited to the softest
energies (1--5 keV). The observations of 4U1608-52 suggest that these QPOs are
present only when the source X-ray luminosity  is within
a rather narrow range and they disappear after X-ray bursts. 
Approximately at the
same level of the source luminosity, type I X--ray bursts cease to exist. 

Judging from this complex of properties, we speculate that
a special mode of nuclear burning at the neutron star surface is
responsible for the observed flux variations. Alternatively, some instabilities
in the accretion disk may be responsible for these QPOs.

   \end{abstract}

%

\sloppypar

\section{Introduction}
Since the discovery of the quasi-periodic oscillations (QPO) in the X--ray
flux from GX 5-1 (\cite{vdk85}, Lewin, van Paradijs \& van der Klis
1988), they were considered an 
important probe of the inner part of an accretion disk and a region where the
accretion disk is interacting with a neutron star surface or
magnetosphere (e.g \cite{alpar85}, for a review of the present status of
QPO observations and theoretical models see e.g. \cite{vdk00}). The
characteristic timescales in these regions are short and the QPOs are
usually observed (and expected) at frequencies of tens or even thousands
of Hz. Aperiodic variability is also present at much lower frequencies
(e.g. $10^{-2}$--$10^{-3}$ Hz). In particular, when the spectrum of the 
accreting neutron
star binaries is soft above $\sim$ 5 keV (e.g. during the so-called
atoll banana state) a Very Low Frequency Noise (VLFN) is
observed at these frequencies (e.g. \cite{hasingervdk89}), with
an approximately power law dependence of power on frequency and
typical RMS variations at the level of percents. Using RXTE data on
several accreting neutron star 
binaries, we searched for low frequency quasi--periodic variations of
the X--ray flux and found clear signatures of QPOs with a surprisingly
similar frequency in 3 sources. The properties of the newly-found
low frequency QPOs are very distinct from those of ``canonical'' high
frequency QPOs, probably indicating a different underlying
physics. 

We describe the experimental results in Section 2. In Section 3 we
speculate on a possible origin of the low frequency QPOs. Section 4
summarizes our findings.

\section{Observations, data analysis and results}
\begin{figure}[htb]
\epsfxsize 8cm
\epsffile[25 185 550 720]{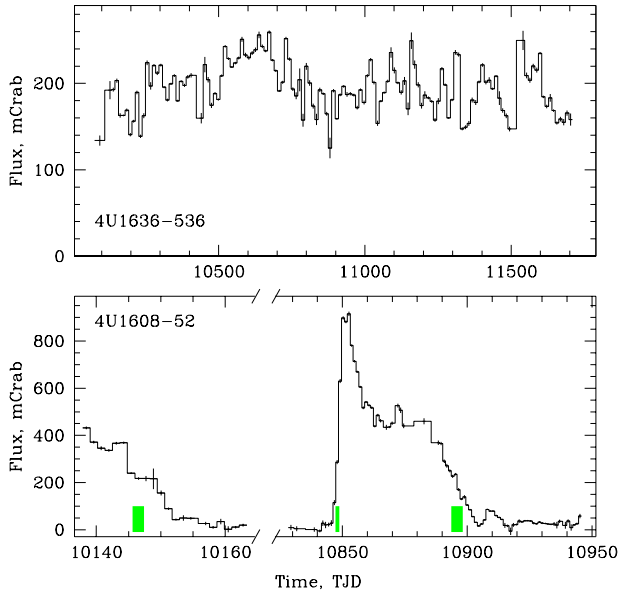}
\caption{Light curves of 4U1608-52 and 4U1636-536 according to
ASM/RXTE data. The gray boxes on the lower panel (for 4U1608-52) mark
the dates of mHz QPO detections.
\label{asm_lcurve}}
\end{figure}

In our analysis we used publicly available data of the Rossi X-Ray Timing
Explorer and EXOSAT observatories. 
We analyzed $\sim$450 ksec of RXTE/PCA observations of 4U1608-52 and $\sim$400
ksec of RXTE/PCA observations of 4U1636-536 covering the period
Mar.1996--Feb.1999. Also, we reanalyzed an archive EXOSAT/ME observation
of Aug.8, 1985.  

Both considered X-ray sources 4U1608-52 and 4U1636-536 are neutron
star binaries which demonstrate type I X-ray bursts. Below we assume a
distance of 4 kpc for 4U1608-52 (e.g. \cite{gottwald87}) and 5 kpc for 
4U1636-536  (e.g. \cite{lawrence83,inoue84}). 4U1636-536 is a
relatively stable  source with an average flux of approximately $\sim$200
mCrab and luminosity $\sim10^{37}$ ergs/s (see Fig.\ref{asm_lcurve}). 
On the contrary, 4U1608-52 is 
a transient source, changing from a quiescent state ($L_{\rm x}<10^{33}$
ergs/s, \cite{asai96}) to a high state with the X-ray flux over
$10^{37}$ ergs/s (e.g. \cite{gottwald87}). Long-term light curves of
the two sources (RXTE/ASM data) are shown in Fig. \ref{asm_lcurve}.

The analysis of the RXTE data was performed with the help of the
standard FTOOLS 5.0 package. For the construction of power
density spectra, the light curves were cleared out of type I
X-ray bursts.

\begin{figure}[htb]
\epsfxsize 8cm
\epsffile[25 185 550 600]{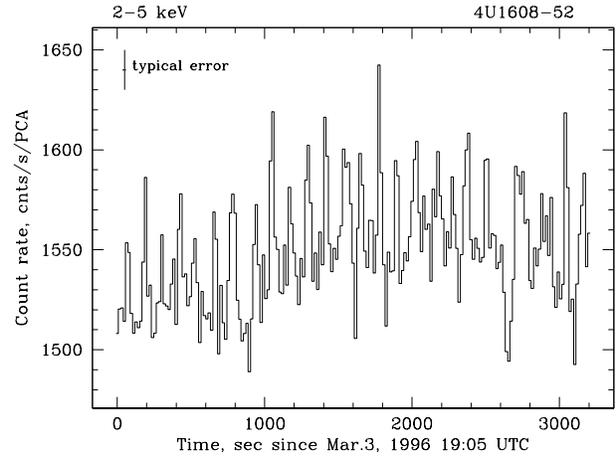}
\caption{A segment of a light curve of 4U1608-52 with QPO. Quasi-periodic 
oscillations are clearly seen.
\label{lcurve}}
\end{figure}

\begin{table*}
\caption{ Parameters of mHz QPOs in the power spectra of 4U1608-52 and
4U1636-536\label{pds_params}}
\tabcolsep=0.32cm
\begin{tabular}{lcccc}
\hline
&\multicolumn{2}{c}{4U1608-52}&\multicolumn{2}{c}{4U1636-536}\\
\hline\\
Observatory/Instrument&RXTE/PCA&RXTE/PCA&RXTE/PCA&EXOSAT/ME\\
Energy band&$\sim$2-5 keV&$\sim$2-5 keV&$\sim$2-5 keV&$\sim$0.8-3.6 keV\\
Date&Mar.3--6, 1996&Mar.24--27, 1998&Apr.1996--Feb.1999&Aug.8,1985\\
Luminosity (3--20 keV), $10^{37}$ erg/s&0.7--1.1&0.5--0.8&0.7-1.0&0.6\\
Frequency, mHz&$7.5\pm0.2$&$7.6\pm0.3$&$7.8\pm0.2$&$8.8\pm0.1$\\
Width, mHz&$2.6\pm0.7$&$1.7\pm0.8$&$1.9\pm0.5$&$2.0\pm0.4$\\
Rms Ampl., \%&$1.85\pm0.2$&$1.17\pm0.2$&$0.69\pm0.08$&$0.9\pm0.1$\\
\hline\\
\end{tabular}
\\
\end{table*}

In Fig. \ref{lcurve} we present a small segment of the analyzed light
curve of 4U1608-52 in the low energy spectral band of RXTE/PCA
($\sim$2--5 keV)  that clearly demonstrates 
quasi-periodic oscillations. We clearly detected
similar oscillations of X-ray flux from 4U1608-52  
during two periods: during the decay phase of X-ray flares of the source in
March 1996 and March 1998. Indications for quasi-periodic variations of
the X--ray flux were also found during the rise phase of the 1998 X-ray
flare (in Feb.3, 1998 and from Mar.14, 1998 till Mar.30, 1998).
But a strong, prominent QPO peak on the power spectra was visible only in
the Mar.3--6, 1996 and Mar.24--Mar.27, 1998 observations. 
Therefore, we use  only these data in the subsequent analysis.
The dates of QPO detection \footnote{including the period during the
rise  phase of the X--ray
flare in Spring 1998, when some indications for QPOs were
also found} are shown by gray boxes in Fig.\ref{asm_lcurve}. 
Neither before nor after these episodes
the QPO with similar frequency and width was detected with an 
approximate 2$\sigma$ upper limit of $\sim$0.4--0.5 \% in the $\sim$2--5 keV
energy band. We note here that the absence of QPOs at lower source
flux levels is not due to  statistical limitations. Similar QPOs with
the RMS of $\sim$1\% would be detectable down to very low fluxes
(down to $\sim$ 100-200 cnts/s or $\sim$10--20 mCrab, i.e. to
luminosity $\sim 10^{36}$ ergs/s).

\begin{figure}[htb]
\epsfxsize 8cm
\epsffile[85 185 460 700]{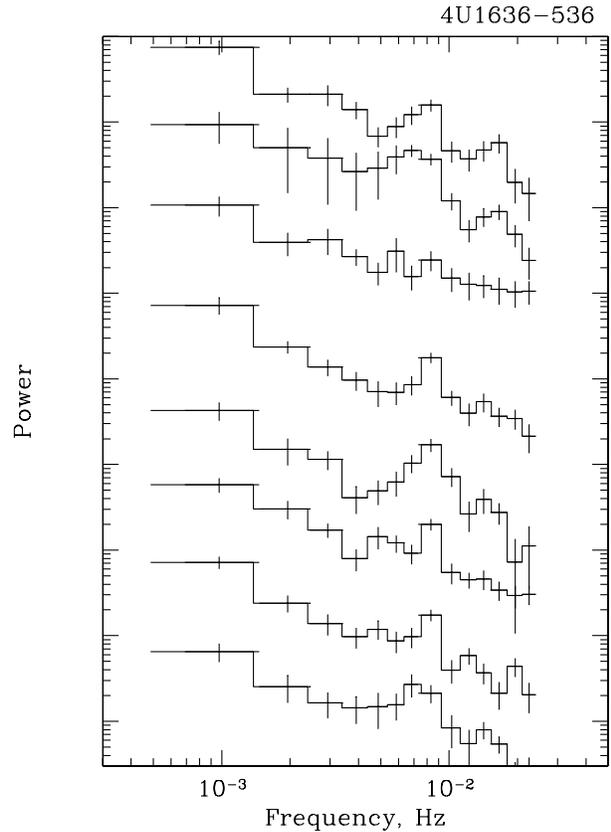}
\caption{The power spectra of 4U1636-536 according to data of RXTE/PCA
observations from 1996 to 1999. Each power spectrum represents
approximately $\sim$50 ksec of observations. The power
spectra of different observational sets were rescaled for clarity.
It is seen that the centroid frequency of the weak QPO peak remains stable. 
\label{setofpowers}}
\end{figure}

On the contrary, quasi-periodic variations are almost always present
in the X-ray flux from 4U1636-536 . Unfortunately,
the QPO in this source is weaker and it
is not always possible to detect the QPO  with sufficient significance
during a single observation. In
Fig. \ref{setofpowers} we plot the power spectra of 
4U1636-536 in several observational sets. It is seen
that a weak QPO peak is almost always present in the power
spectrum. Moreover, the centroid frequency of the QPO is very stable --
$\sim$7--9 mHz. In the subsequent analysis we used the power
spectrum of 4U1636-536 obtained by averaging over all observations
from Mar.96--Feb.99.  

\begin{figure*}[htb]
\hbox{
\epsfxsize 5.8cm
\epsffile[10 185 550 720]{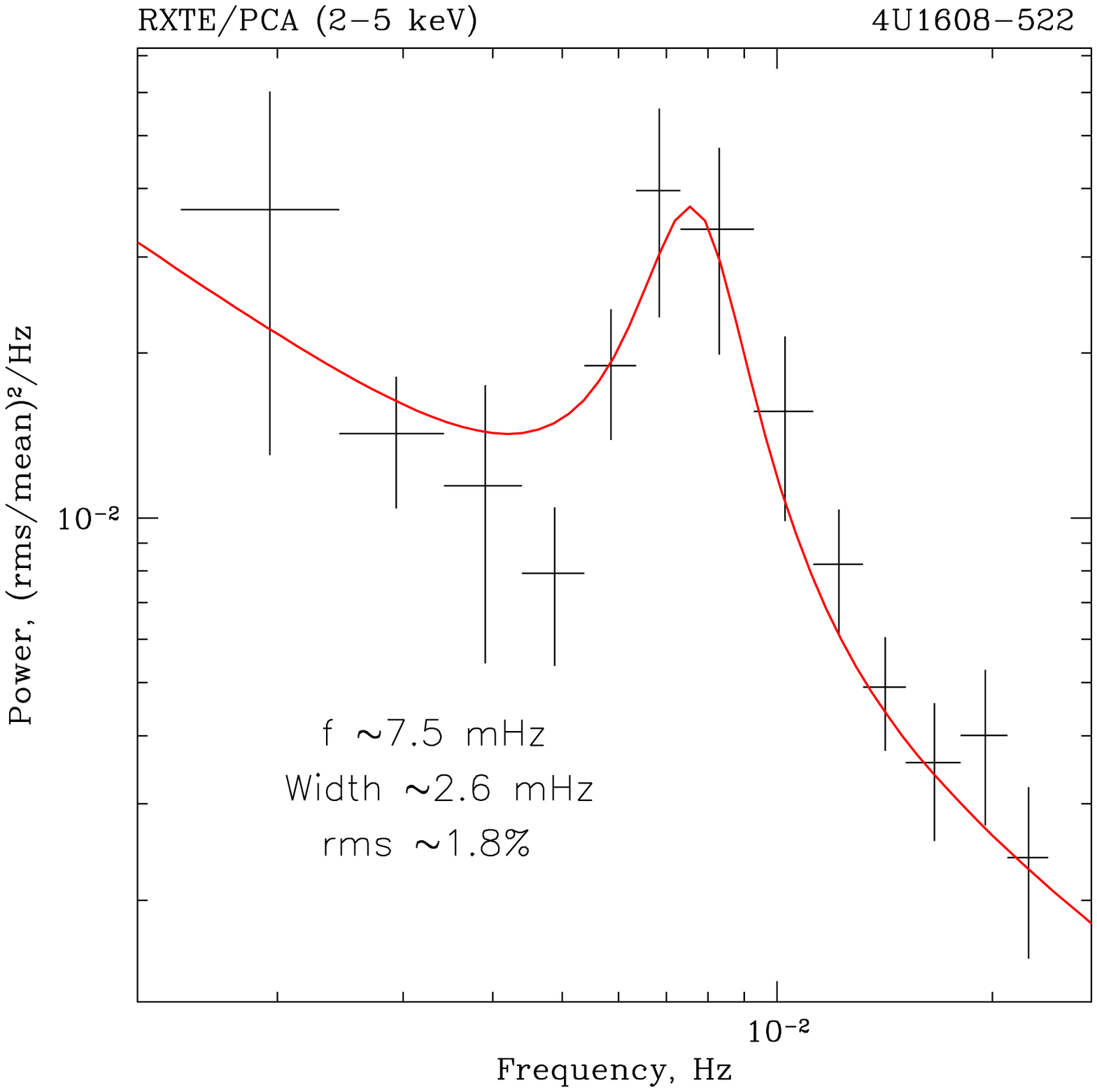}
\epsfxsize 5.8cm
\epsffile[10 185 550 720]{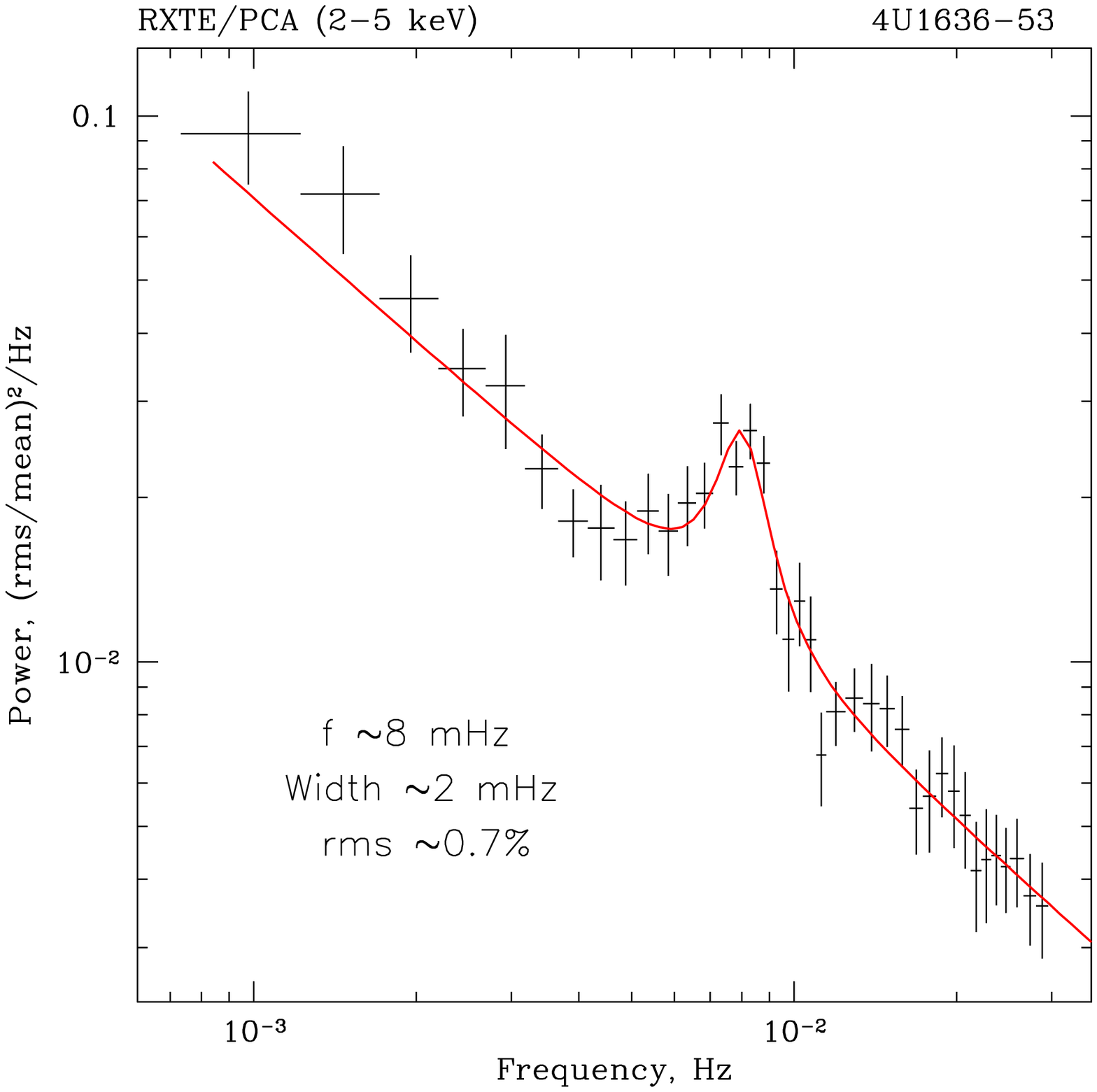}
\epsfxsize 5.8cm
\epsffile[10 185 550 720]{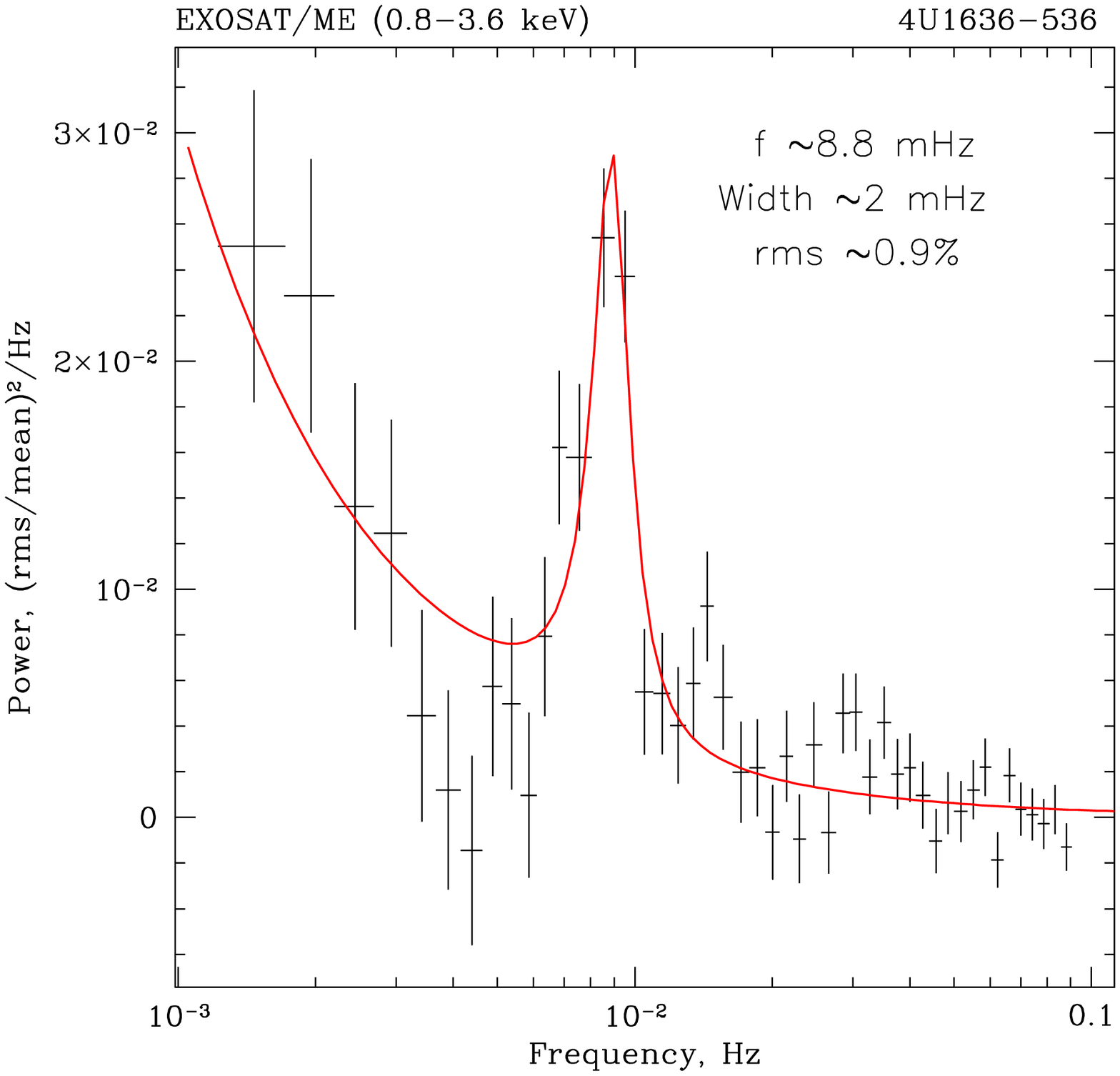}
}
\caption{The power spectra of 4U1608-52 and 4U1636-536 with mHz
QPO. {\it Left panel:} 4U1608-52, observations Mar.3--6, 1996; {\it
middle panel:}4U1636-536, observations Mar.1996--Feb.1999; {\it right
panel:} 4U1636-536, archive observations of EXOSAT/ME, Aug.8, 1985.
\label{powers}}
\end{figure*}

We then analyzed the archive data of the EXOSAT/ME observation of 4U1636-536
on Aug.8, 1985. A weak QPO at a frequency of $\sim$9 mHz  was also
found (see Table 1 and Fig.\ref{powers}). A similar search for the low
frequency QPO in the archival EXOSAT/ME observations of 4U1608-52
also revealed a possible 10 mHz QPO candidate (though weak) during
an observation on July 5, 1984.

The power density spectra (PDS) for the 2--5 keV light curves of 4U1636-536
and 4U1608-52 were constructed in the $\sim 10^{-3}$--0.03 Hz frequency range. The PDS
were fitted with a model consisting of a power law (VLFN) and a
Lorentzian (QPO). The parameters of the detected QPOs in the light
curves of 4U1608-52 and 4U1636-536 are presented in Table
\ref{pds_params}. The power spectra of 4U1608-52 and 4U1636-536 along
with the best fit models are shown in Fig. \ref{powers}. 

\begin{figure}[htb]
\epsfxsize 8cm
\epsffile[25 185 550 720]{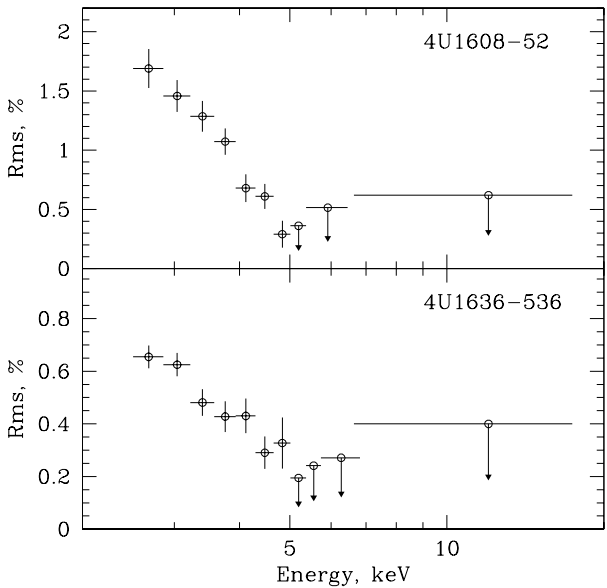}
\caption{The dependence of rms amplitude of QPO on the photon
energy. For 4U1608-52 the observations  from March 1996 was
used. 2$\sigma$ upper limits are shown for energies greater than 5 keV.
\label{rms_en}}
\end{figure}

In order to derive the energy dependence of the QPO amplitude, 
we constructed power density spectra in each
energy channel of PCA. These power density spectra were fitted with
the same model, fixing the slope of the VLFN component, the centroid
and width of the QPO peak. The resulting dependencies of the amplitude 
of mHz QPOs on the photon energy are presented in Fig.\ref{rms_en}.

\section{Discussion}
The properties of the observed mHz QPOs (mQPOs hereafter) can be
summarized as follows:
\begin{itemize}
\item The fractional RMS amplitude strongly decreases with energy
\item Flux variations in mQPOs are at the level of percents
\item mQPOs seem to be present only at a particular level of source 
X-ray luminosity: $L_{\rm x}\sim$0.5--1.5 $\cdot
10^{37}$ ergs/s. However this tentative conclusion is based on
the data on only two sources and more observations are need to
confirm it.
\item mHz QPOs ($f\sim$ 7--9 $\cdot 10^{-3}$ Hz)  with similar properties are
found in two ``atoll'' sources
\item mQPOs with similar centroid frequencies are present in data for 
4U1608-52 and 4U1636-536 that are separated by years (or even $\sim$14
years for 4U1636-536).
\end{itemize}
The decline of the fractional RMS amplitude of a QPO with energy
(Fig.\ref{rms_en}) 
seems to be a rather unusual property. Typically, the fractional RMS
amplitude of the variability (at least at frequencies $\sim$ Hz or
higher) increases with energy for accreting NS LMXBs(see 
e.g. \cite{lewin92,berger96,zhang96} Revnivtsev, Borozdin \& Emelyanov 
1999, \cite{homan99}). The same is true for the Very 
Low Frequency Noise (VLFN) in 4U1608-52 and 4U1636-536 at
frequencies comparable to the mQPO frequency. This unusual
dependence of RMS on energy possibly hints at a very different nature of the 
flux variations in mQPOs. Below we propose several possible
explanations for the observed phenomenon.

\subsection{Quasi-periodic nuclear burning?}
The possibility that at high mass accretion rates nuclear
burning may cause low frequency luminosity variations was first
suggested by 
Bildsten 1993,1995 (for an overview of the nuclear burning regimes see
e.g. Lewin, van Paradijs, Taam 1993, Bildsten 1997,2000). Although
the original suggestion of Bildsten (1993) was that VLFN is a
signature of nuclear burning, we speculate below that the newly 
discovered mHz QPOs may correspond to some special mode of nuclear burning
which only occurs in a certain range of the mass accretion rate.

First of all, the modulation of the flux at the level of percents (see
Table 1 and Fig.\ref{powers}) is roughly consistent with the expected relative
energetics of accretion and nuclear burning ($L_{\rm acc}/L_{\rm
nuc}$ of the order of 100). For the March 3, 1996 observation of
4U1608-52 the ratio 
of the total source luminosity to the luminosity of the variable
component, averaged over the QPO period, is estimated as
$L_{\rm tot}/L_{\rm mQPO}\sim 145$. 

\begin{figure}[htb]
\epsfxsize 8cm
\epsffile[25 185 550 620]{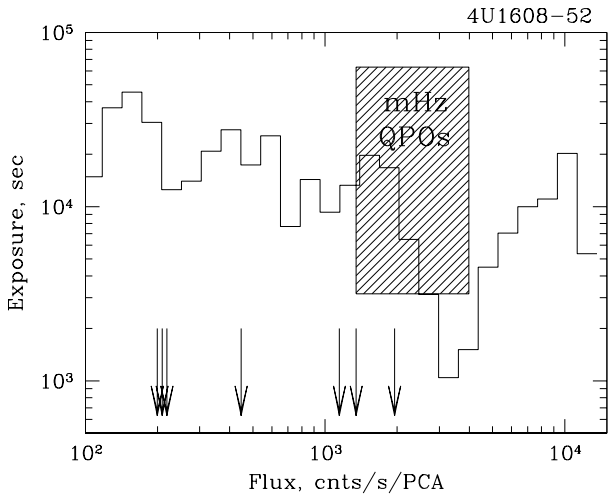}
\caption{The PCA/RXTE exposure time for 4U1608-52 at different flux
levels. Shaded region shows the range of fluxes when mQPOs were
observed. Arrows mark persistent fluxes of the source when Type I
bursts were observed. 
\label{cartoon}}
\end{figure}

Secondly, it seems that mHz QPOs in 4U1608-52 were present only
when the source flux was in a narrow range (within a
factor of 2-3), while the total range of the source flux variations during
the RXTE observations spans two orders of magnitude
(Fig.\ref{asm_lcurve},\ref{cartoon}). Interestingly, at
approximately the same flux level, type I X--ray bursts cease to 
exist, as shown in Fig.\ref{cartoon}. Although the number of bursts
detected is not very large,
the coincidence is striking and may hint at an intimate relation
between changes in the nuclear burning regimes and observed 
QPOs. This assumption, that mHz QPOs are associated with a specific
range of mass accretion rates, implies that 4U1636-536, whose flux did
not change much during the RXTE observations, by chance has about the right
accretion rate, for generation of mHz QPOs.

\begin{figure}[htb]
\epsfxsize 8cm
\epsffile[25 185 550 620]{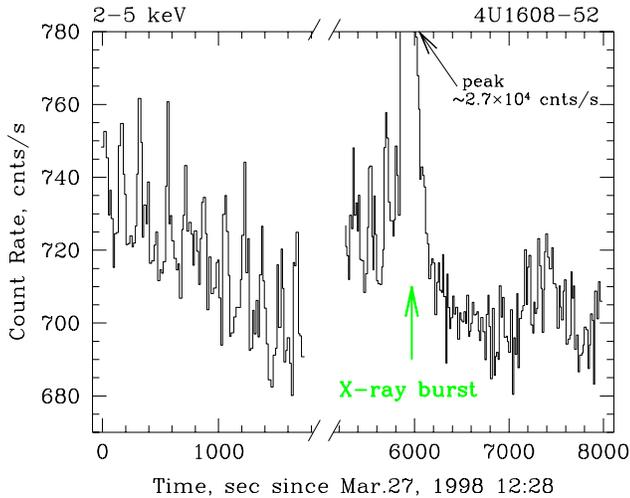}
\caption{The light curve of 4U1608-52 before and after X-ray burst of
4U1608-52 on Mar.27, 1998. Note that quasi-periodic variations are
clearly visible prior to the burst, but cease after the burst.
\label{burst}} 
\end{figure}

Further support for the possible link between nuclear burning and the mHz
QPOs in 4U1608-52 comes from the analysis of the light curve before
and after a type I X--ray burst that occurred during the period when
the mHz QPOs were detected.  The relevant part of the 2--5
keV light curve is shown in Fig.\ref{burst}. Quasi-periodic
oscillations are clearly visible during the orbit
preceding the one in which the X--ray burst was detected. Moreover mQPOs are
also present 
immediately before the burst, but cease after the burst. This behavior
would naturally fit the assumption that nuclear burning is responsible
for mQPOs, if a large fraction of the fuel is consumed during the type 
I burst and a long time is needed to restore the conditions. Note here
that for 4U1608-52 the ratio of peak X-ray fluxes of a type I 
X-ray burst and small ``microburst'' is of the order of 600-700.
A ratio of the total energies released in the single type I burst and in
the variable component during one $\sim$ 120 s period of mQPO is
of the order of 150-180.  The latter number may indicate that compared to 
a typical type I X--ray burst, the amount of  nuclear fuel consumed during 
a single cycle of the mQPO is less than per cent.

A change in the  X-ray flux variability after a Type I X-ray burst
was previously reported by Yu et al. (1999) for Aquila X-1. The authors   
reported a drop in the VLFN level after the burst that was accompanied 
by changes in the source flux and a decrease in the kHz QPO
frequency. Possible correlation of the low frequency variability with
nuclear burning at the neutron  
star surface was also mentioned. We reanalyzed the archival data of
the observation of Aql X-1 described in Yu et al. 1999 (Mar.1, 1997) and
found a  QPO peak at a frequency of
$\sim$6--7 mHz. These variations also have a soft spectrum and are
undetectable at the energies above $\sim$ 7 keV.
 Moreover, a significant fraction of the VLFN at these frequencies can
be attributed to this QPO. Again, as in the case of  
4U1608-52,  this QPO--like feature
becomes undetectable after the burst. Thus, although  it is difficult
to offer a satisfactory
explanation of all changes occurring during the burst, we
conclude that the disappearance of mQPOs is consistent with their
``nuclear'' origin. 

\begin{figure}[htb]
\epsfxsize 8cm
\epsffile[25 185 550 720]{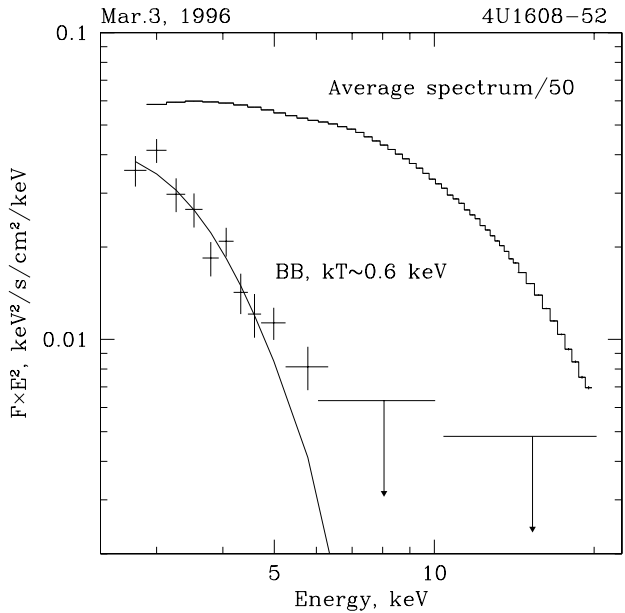}
\caption{The energy spectrum of the variable component in 4U1608-52. This
spectrum was obtained using the integrated rms value of mHz QPO variability 
of X-ray flux in different energy channels (so called frequency resolved 
spectral technique, see e.g. \cite{freq_res}). The solid line represents the 
approximation of this spectrum by a black body model with temperature $kT=0.6$
 keV. For comparison the averaged
spectrum of the source is shown (scaled down by a factor of 50 for
easier comparison with the variable component).
\label{spec}}
\end{figure}

Next, we examine the spectrum of the ``variable'' component of the mQPO in
4U1608-52. The spectrum was constructed using the integrated rms value
 of mHz QPO variations of
 the X-ray flux in different energy channels, i.e. using the so-called frequency 
resolved spectral technique (see \cite{freq_res}). The resulting spectrum 
is shown
in Fig.\ref{spec}. \footnote{Note that this spectrum characterizes the
amplitude of flux variations  as a function of energy and
therefore does not necessarily correspond to the existing
spectral component.} For comparison, in the same plot, the averaged
spectrum of the source is shown (scaled down by a factor of 50 to
facilitate comparison with the variable component). It is obvious that the
``variable'' 
spectrum is much softer than the averaged spectrum. This was of course
expected given the strong decline of the QPO rms with energy
(Fig.\ref{rms_en}). For comparison, the solid curve shows  black body
emission with a temperature of $\sim 0.6$ keV\footnote{We note here that
a very soft component was also found by Yu et al. 1999 by
subtracting the spectrum accumulated after the type I burst from the spectrum
observed before the burst in Aql X-1}. A detailed fit of the soft
component is difficult to make because of the poor statistics. It is however
certain that the spectrum is softer than the $\sim$1.5--2 keV black body,
characteristic of type I X--ray bursts. The total luminosity
of the source during this observation was $\sim
10^{37}$ erg~s$^{-1}$. If the few percent modulations are caused by 
quasi-periodic nuclear burning over the whole neutron star surface, then the
corresponding temperature will be $T\sim (fL/4\pi\sigma
r^2)^{1/4}\sim$ 0.3--0.5 keV, where $f$ is the ratio of the peak nuclear energy
release to the accretion luminosity, $r\sim$8--10~km is the neutron star
radius, $\sigma$ is the Stefan--Boltzmann constant. This estimate
assumes that no strong sources of heating other than nuclear burning are
present. I.e. it assumes that a significant part of the burning take place
outside the region where accretion energy is released  and the
surface of the neutron star is ``preheated'' to a significantly larger
temperature. In the picture of a spreading layer on the surface of a neutron
star (\cite{inogamov99}), the kinetic energy of the accreting flow is
released in two bright rings equidistant from the equator. At luminosities
of the order of  
$10^{37}$ ergs/s these rings occupy only $\sim$15--20\% of the stellar
surface. Nuclear burning could occur in any part of the star surface
but in much deeper layers and column densities.
Variations of the nuclear burning energy release in the
region where the bulk of the accretion energy is released would
result in variations of the much harder component which dominates
the averaged spectrum of the source. These variations could be
responsible for a possible weak harder component that in principle could be 
under the 2-$\sigma$ upper limits presented in Fig.\ref{spec} as the extension of 
the soft component at energies
higher than 6 keV. If this interpretation is correct, then the weakness of
the harder component implies that only a small fraction of burning 
($\le$ 20\%) occurs in the area where accretion energy is
released.

\begin{figure}[htb]
\epsfxsize 8cm
\epsffile[25 185 550 580]{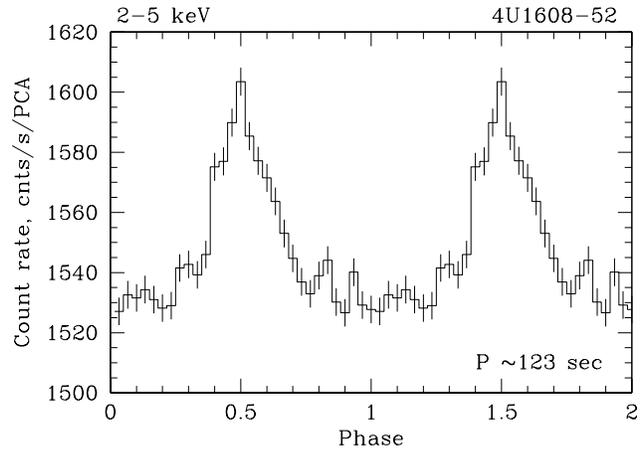}
\caption{The profile of the mHz QPO, obtained by folding the light
curve of 4U1608-52 obtained during the first orbit of observation on
Mar.3, 1996. The error bars shown are due to pure Poissonian noise.
\label{pulse}}
\end{figure}

Finally, we examined the characteristic shape of the mQPO in 4U1608-52
using the portion of the light curve where individual QPO profiles are
clearly visible. The light curve was divided into pieces with the
duration of $(8.13 \cdot 10^{-3}$~Hz$)^{-1}\sim 123$~s. The phase of the
profile in each piece was then calculated by comparing the observed
profile with a sine wave, and the profiles were coadded with the
proper shift of phase. The resulting QPO profile is shown in
Fig.\ref{pulse}. Our experiments with various template profiles to 
determining the phases of individual pulses resulted in moderate
variations in the averaged pulse profile (as expected, given the
limited statistics of the data set). The characteristic features of
the profile, namely the narrower peaks  and more extended valleys, are
however robust against almost any choice of the procedure of 
constructing the averaged profile. There is also marginal evidence
that the peaks are asymmetric: the steeper rise and shallower decline.  It is
interesting that this profile bears some similarity to the time
dependent energy release plots of Bildsten 1995 (see his Figure
5a), which were calculated using the approximation of a
one-dimensional none-convective model.  
However, the mass accretion rate (the Eddington level) assumed for these plots
 is markedly different from what we consider here.

Thus, there are at least several observational facts that seem to
be consistent with  association of the mQPOs with the
quasi-periodic  burning of the nuclear matter at the
neutron star surface. It is unclear, however, what process governs the
time scale of $\sim$ 2 minutes, which is surprisingly stable (and even
similar for two different sources), and why the burning is quasi--periodic
and present only in a narrow range of mass accretion
rates.

If the nuclear burning is not uniform over the neutron star surface
one may expect asymmetries in the distribution of hot spots. This
might lead to the appearance of coherent pulsations with 
the neutron star rotation period, as is indeed observed
during type I X-ray bursts (see e.g. \cite{stroh}). 
 We tried to search the 4U1636-536 data (the type I X-ray bursts
excluded) for coherent pulsations at the pulse frequency 580 Hz
(e.g. \cite{stroh98}). We made
no solar baricentric or pulsar orbital corrections. In order to
take into account these possible changes of observed pulse frequency we
took a 1 Hz wide detection cell. We obtained an upper limit on the
modulations in the coherent signal of $\le $0.2 \% (95\% confidence)
pulse fraction for the 2--5 keV range. This result sets a limit on
the uniformity of energy release on the neutron star
surface. This limit is, however, not very stringent given the small $\sim$ 0.7\%
(see Table 1) contribution of nuclear burning to the source luminosity.

Trying to undestand the nature of unstable nuclear burning we see two
simplest possible modes:

\begin{itemize}
\item  unsuccessful ignitions occur from time to time in different parts 
of the freshly accreted fuel. A flame  burns only on a
small surface area. The whole stock of fuel is waiting for a successful flame
front capable of propagating across the entire surface of the star. 
    It is obvious that such a picture could not produce quasiperiodic
oscillations. The resulting variability should be much more stochastic 
and could
produce only broad band noise.

\item unstable helium shell burning occurs at some depth in the
freshly accreted fuel. The flame propagates over the whole surface of
the star. In this case it is possible to expect quasiperiodicity.
However, the observation of a normal type I X-Ray burst after a long
series of microbursts (see Fig. 7)  strongly restricts possible models of
shell burning. The most important consequence is that unstable shell
burning occurs below the main fuel stock and does not influence it
until the conditions for the strong burst materialise. In addition, this
picture requires that only a small fraction of the fuel is processed during the
shell burning, leaving enough fuel for a stronger type I burst. Questions
about flame propagation through a thin shell are not simple either. 
\end{itemize}

If the unstable nuclear burning interpretation is correct then one can
make two obvious predictions:
\begin{itemize}
\item Other NS LMXB may show similar QPOs while being at a certain
flux level. For 4U1608-52 and 4U1636-536 the
3--20 keV luminosity was $\sim$ 0.5--1.5 $\cdot 10^{37}$~ergs/s when
the mQPOs were observed.
\item Low frequency QPOs with similar properties (in particular
with a much softer spectrum than the average one) should not be present in
X-rays from black hole candidates. 
\end{itemize}

\subsection{Disk instabilities}

Although there are several observational indications that mQPOs may be
related to nuclear burning on the surface of the neutron star, the
observations do not provide robust enough proof of this
interpretation. We briefly discuss below a few other  scenarios.

The soft spectrum of the variable component (much softer than the
averaged spectrum) may be hinting at a possible contribution of an
optically thick accretion disk emission to the variable component.
In this case we can also expect that a powerful type I X-ray burst
could lead to the disappearance of disk instabilities for some period.

We note here that  variations of the soft
component on the time scales of 100-1000 s have been observed in 
the galactic black hole candidate GRS 1915+105
(\cite{belloni97}, Trudolyubov, Churazov \& Gilfanov 1999,
\cite{muno99}), which are thought to be associated 
with the motion of the inner boundary of an optically
thick accretion disk.  It is  not clear however if these
variations have the same nature as the mQPOs discussed 
above. These variations appear in GRS 1915+105 when the average
source luminosity is close to the Eddington limit, while the QPO
discussed above were observed in sources with  luminosities of only
$\sim0.1  L_{\rm edd}$. The amplitude of variations observed in
GRS 1915+105 is very large - more than an order of magnitude. 

\subsection{Other possible scenarios}

It is interesting that variations of the photoelectric absorption at
the level of $N_{\rm H}L\sim 3 \cdot 10^{21}$~cm$^{-2}$ are capable
of reproducing approximately the required dependence of the RMS on
energy. It is not clear however what kind of process can cause
quasi-periodic variations of the absorption on the time scales of
minutes, especially given the stability of the mQPO frequency.

Variations of mass accretion rate in an optically thick accretion
disk may also be responsible for the observed variations. However the
amplitude of variations of the harder component, presumably coming
from the boundary layer, is at least a factor of
5 lower than that for the soft component. This makes the variations in  the
mass accretion rate an unlikely cause of the observed mQPOs.

\section{Conclusions}
A new type of very low frequency QPO  has
been discovered in at least  two ``atoll'' sources: 4U1608-52 and
4U1636-536, with essentially the same frequency $f\sim$ 7--9 $\cdot 10^{-3}$
Hz. These QPOs have a very soft spectrum, RMS amplitude of variations
of the order of per cent and appears to be present only 
at a certain level of the mass accretion rate. We suggest that these
quasi--periodic variations might be related to a special regime of
the nuclear burning on the neutron star surface or some instability in
the optically thick accretion disk.

\begin{acknowledgements}
We are grateful to the referee Michael van der Klis for useful comments.
This research has made use of data obtained 
through the High Energy Astrophysics Science Archive Research Center
Online Service, provided by the NASA/Goddard Space Flight Center.
M.Revnivtsev acknowedges partial support by RFBR grant 00-15-96649. 
\end{acknowledgements}

\end{document}